\documentclass[9pt,lineno]{elife}

\usepackage{lipsum} 
\usepackage[version=4]{mhchem}
\usepackage{siunitx}
\DeclareSIUnit\Molar{M}

\title{A Hypothesis-First Framework for Mechanistic Modeling in Neuroimaging}
\DeclareFloatingEnvironment[name={Supplemental Figure}]{suppfigure}
\author[1*]{Dominic Boutet}
\author[1*]{Sylvain Baillet}
\affil[1]{Montreal Neurological Institute, McGill University, Canada}

\corr{dominic.boutet@mail.mcgill.ca}{DB}
\corr{sylvain.baillet@mcgill.ca}{SB}

\presentadd[\authfn{3}]{McConnell Brain Imaging Centre at Montreal Neurological Institute, McGill University, Canada}
\presentadd[\authfn{4}]{McConnell Brain Imaging Centre, Montreal Neurological Institute, McGill University, Montreal, QC H3A 2B4, Canada}

\begin{document}
	
	\maketitle
	
	\begin{abstract} 
		Turning rich neuroimaging data into mechanistic insight remains challenging. Statistical models capture associations but remain largely agnostic to underlying mechanisms. Biophysical models embody candidate mechanisms but remain difficult to deploy without specialized expertise. Here, we present a hypothesis-first framework recasting model specifications as testable mechanistic hypotheses and streamlines the procedure for rejecting inappropriate hypotheses \textit{before} moving to typical analyses. The key innovation is an expectation of model behavior under feature generalization constraints: we compute the model's expected $Y$ output across the parameter space based on the likelihood for a broader/distinct feature $Z$. Mirror statistical models are derived from these expected outputs and compared to the empirical ones with standard statistics. In synthetic experiments, our framework rejected mis-specified hypotheses and penalized unnecessary degrees of freedom while retaining valid hypotheses. These results demonstrate a practical hypothesis-driven approach for using mechanistic models in neuroimaging without requiring advanced training, complementing traditional analyses. 
	\end{abstract}

	\section{Introduction}
	
	Neuroimaging provides rich, multiscale measurements of brain structure and activity, but translating these measurements into mechanistic understanding remains a major challenge. Statistical models are the dominant analytical tools in the field: they capture significant relationships between variables and quantify their strength, but they are typically agnostic about the biological or dynamical processes that generate those relationships \citep{friston_computational_2010, mcintosh_multivariate_2013}. Mechanistic models, in contrast, explicitly instantiate candidate biological processes, often as dynamical systems, and can therefore offer causal explanations and predictive power that statistical approaches alone cannot \citep{kirk_model_2013,breakspear_dynamic_2017,shine_computational_2021}.
	
	However, the practical adoption of mechanistic modeling in neuroimaging remains limited. These models are powerful but require numerous technical decisions, such as model form, parameterization, priors, and constraints; each of which can strongly influence results. Without specialized training, it is easy to overlook issues such as parameter identifiability, non-physiological model regimes, or the consequences of under- or over-parameterization \citep{kirk_model_2013,heinrich_structural_2025}. As a result, mechanistic modeling remains the domain of a relatively small subset of researchers, even though its potential to generate biological insight is widely recognized.
	
	Here, we propose a hypothesis-first framework designed to lower these barriers and integrate mechanistic modeling more directly into standard neuroimaging workflows. The central idea is to treat modeling choices, including the equations used, which parameters are free or fixed, and the structure of their interactions, as explicit, testable mechanistic hypotheses. Instead of attempting to infer parameters immediately, we first evaluate whether a given mechanistic hypothesis is generally appropriate for capturing a statistical relationship of interest. This allows researchers to reject ill-posed hypotheses before looking at parameter estimation or simulation-based inference results, thereby ensuring that downstream analyses are conducted on models that instantiate an appropriate mechanistic hypothesis.
	
	The framework introduces two key innovations. First, it incorporates a novel approach to incorporating feature generalization constraints: rather than evaluating a mechanistic model solely by how well it reproduces a single target feature of interest $Y$ (e.g., alpha power in MEG data) at its optimal parameter set, we evaluate an expectation on the model's reproduction of $Y$ computed using a likelihood defined by how well it reproduces a broader or distinct feature $Z$ (e.g., the full power spectrum) throughout the defined parameter space. This discourages non-physiological regimes that over-fit $Y$ at the cost of respecting the broader or distinct feature $Z$  and ensures that the explanatory power of the downstream models later derived using parameter estimation/inference will generalize beyond a single metric. Second, we construct a mirror dataset from these expectations in model $Y$ outputs which is used to obtain a mirror statistical model and compare it directly to the empirical statistical model. This comparison uses pre-defined standard statistical tests, such as tests comparing model coefficients and residuals distributions, providing straightforward accept/reject criteria for each hypothesis based on the specific objectives of the empirical statistical relationship of interest. Thus, a rigorous implementation of mechanistic modeling withing our framework depends on domain knowledge for defining mechanistic hypotheses along with feature generalization settings instead of specialized technical training. Making mechanistic modeling more accessible to neuroimaging researchers with expert domain knowledge.
	
	We demonstrate the utility of this framework using synthetic datasets with known ground truth generated from the Wilson-Cowan neural mass model. These experiments show that the framework can correctly reject hypotheses which associated models are under-parameterized, overly constrained, or reliant on invalid assumptions; penalize over-parameterized models; and retain appropriately specified models. By grounding mechanistic modeling to hypothesis formulation and validation, this approach enables a more rigorous, interpretable, and accessible use of mechanistic computational models in neuroimaging, bridging the gap between statistical association and mechanistic explanation.
	
	\section{Methods}
	
	\subsection{A Hypothesis-First Framework for Mechanistic Modeling}
	Our goal is to provide a structured, hypothesis-first procedure that complements standard mechanistic modeling workflows and can be integrated into typical neuroimaging analyses. Here, we briefly describe the conceptual components of the framework, using a hypothetical neuroimaging scenario as an illustrative example. A more formal mathematical treatment is provided in the Supplementary Information.
	
	\subsubsection{\textit{Empirical Data \& Statistical Model}}
	We begin with a conventional statistical analysis, which serves as the empirical foundation for subsequent mechanistic modeling.
	
	Consider a dataset $D=\{{D^0},...,{D^N}\}$ consisting of paired neuroimaging measurements: magnetoencephalography (MEG) signals $X^n$ and structural MRI data $M^n$. From these raw data, we derive relevant features: for instance, a cortical microstructural index $S=G(M)$ from MRI feature, and a power spectral density (PSD) $\gamma=\lambda(X)$ from MEG. Since PSDs are high-dimensional, we often reduce them to a summary measure such as alpha-band power $Y=F(\gamma)$. We can then model the relationship between between $S$ and $Y$ using a statistical model $R$ (e.g., linear regression), obtaining predicted values $\hat{Y}_R$, regression coefficients $\beta_R$, and residuals $\hat{\epsilon}_R = Y-\hat{Y}_R$. 
	
	At this stage, we have a standard statistical description of the relationship between structure and function, which defines the phenomenon for which we aim to find mechanistic explanations (Figure 1, blue and top green panels) \citep{mcintosh_multivariate_2013,baillet_magnetoencephalography_2017}. 
	
	\subsubsection{\textit{Defining Mechanistic Hypotheses}}
	The next step is to formulate mechanistic hypotheses that could explain the observed statistical relationship. Each hypothesis corresponds to a specific biophysical mechanism, translated into mathematical form as a dynamical model with associated parameterization. Let the set of hypotheses be $\alpha_{\Phi}=\{\alpha_0,...,\alpha_{\phi}\}$. Each hypothesis $\alpha$ defines a computational model $\Psi_{\alpha}$ with parameterization $\theta \in \Theta_{\alpha}$ and optional priors $P_{\alpha}(\theta)$. 
	
	These models take $S$ and $\theta$ as input and generate simulated PSDs $\tilde{\gamma}_{\alpha,\theta} =\Psi_{\alpha}(S,\theta)$. From these, we compute simulated alpha power $\tilde{Y}_{\alpha,\theta} = F(\tilde{\gamma}_{\alpha,\theta})$. For example, one plausible hypothesis might be that microstructural differences modulate the excitatory threshold of cortical populations. This can be implemented by varying the activation threshold parameter in a Wilson–Cowan neural mass model\citep{wilson_mathematical_1973}. 
	
	\subsubsection{\textit{Using Mechanistic Models}}
	At this stage, most mechanistic analyses proceed directly to parameter estimation or inference, either by identifying the parameter set that minimizes a loss function or by inferring posterior distributions over parameters \citep{ritter_virtual_2013,woodman_integrating_2014,calvetti_inverse_2018,wang_inversion_2019,cranmer_frontier_2020,goncalves_training_2020,tejero-cantero_sbi_2020,hashemi_simulation-based_2022,boutet_metaheuristic_2023}. 
	
	Our framework is designed to intervene \textit{before} parameter estimation/inference, providing a systematic way to assess whether a mechanistic hypothesis is generally appropriate for explaining the observed relationship given the data. It allows researchers without extensive training in dynamical systems theory to assess key model properties, such as whether the parameter space supports relevant behaviors or avoids non-physiological regimes, before deeper inference is attempted by relying on their domain knowledge.
	
	For example, if a model's fixed parameters are set so that it cannot generate oscillatory activity under any parameter configuration, then any interpretation of the results derived from it would be misleading. Our framework is explicitly designed to flag such cases from the start.
	
	\subsubsection{\textit{Generating Mirror Statistical Models}}
	The core innovation of our approach lies in two complementary ideas: (1) feature generalization and (2) mirror statistical models reflecting an expectation over the parameter space. 
	
	Feature generalization broadens the evaluation criteria beyond a single feature $Y$. Instead of assessing the model based only on how well it reproduces $Y$, we define an additional feature $Z = T(\gamma)$, often a more comprehensive or distinct property of the system, and compute the expectation of the model output in $Y$ given how well it reproduces $Z$ across the parameter space. This ensures that the model's explanatory power generalizes beyond narrow fits that account for $Y$ exclusively and discourages overfitting or non-physiological solutions. Essentially, we weight each point in the parameter space by how well it reproduces $Z$, then compute the expectation on the model's $Y$ output: 
	\begin{equation}
		E[\tilde{Y}_{\alpha,\theta}|Z] = \int_{\theta \in \Theta_\alpha} \tilde{Y}_{\alpha,\theta} \, P_\alpha(\theta \mid Z) \, d\theta\\
	\end{equation}
	Intuitively, this can be interpreted as asking, \textit{"What does the model predict for $Y$ when we only consider parameter regimes that explain $Z$?"}
	
	We then use this expected output to construct a mirror dataset from which we derive a mirror statistical model $\tilde{R}_\alpha$ for each hypothesis. The mirror model captures the statistical relationship implied by the mechanistic model under feature generalization. By comparing $\tilde{R}_\alpha$ to the empirical model $R$, for example by testing whether their coefficients or residual distributions differ, we obtain simple, binary accept/reject outcomes for each hypothesis.
	
	\subsubsection{\textit{Identifying Inappropriate Mechanistic Hypotheses}}
	The final step is to formalize how mirror models are compared to empirical ones. Standard statistical tests (e.g., $t$-tests, variance tests, distributional comparisons) are used to evaluate whether $\tilde{R}_\alpha$ differs significantly from $R$ in terms of our pre-defined criteria, e.g., differences in coefficients and residuals distributions. Given that the mirror model is derived from a mirror dataset that is meant to reproduce the empirical data, we propose to assess the difference between the empirical statistical model's residuals and the following values: $\bar{\epsilon}_{R_{\alpha}} = Y-\hat{Y}_{R_{\alpha}}$, where $\hat{Y}_{R_{\alpha}}$ corresponds to the predictions of the mirror model $R_{\alpha}$. This amounts to testing whether $R_{\alpha}$ can substitute $R$ as a model of the empirical data, which we deem more appropriate than comparing the residuals distributions derived from their respective datasets. If the mirror model for a given hypothesis fails the statistical tests, the corresponding mechanistic hypothesis is rejected. 
	
	For example, if we have linear regression models and the slope of $\tilde{R}_\alpha$ differs from $R$, this indicates that the hypothesis fails to capture the rate of the statistical relationship under feature generalization constraints. Conversely, finding no difference suggests that the mechanistic hypothesis is appropriate. This procedure yields a binary decision for each hypothesis, enabling researchers to discard inappropriate models before engaging in parameter estimation/inference analyses.
	
	\topfigrule	
	
	\begin{figure}[h!]
		\centering
		\includegraphics[width = 0.95\linewidth]{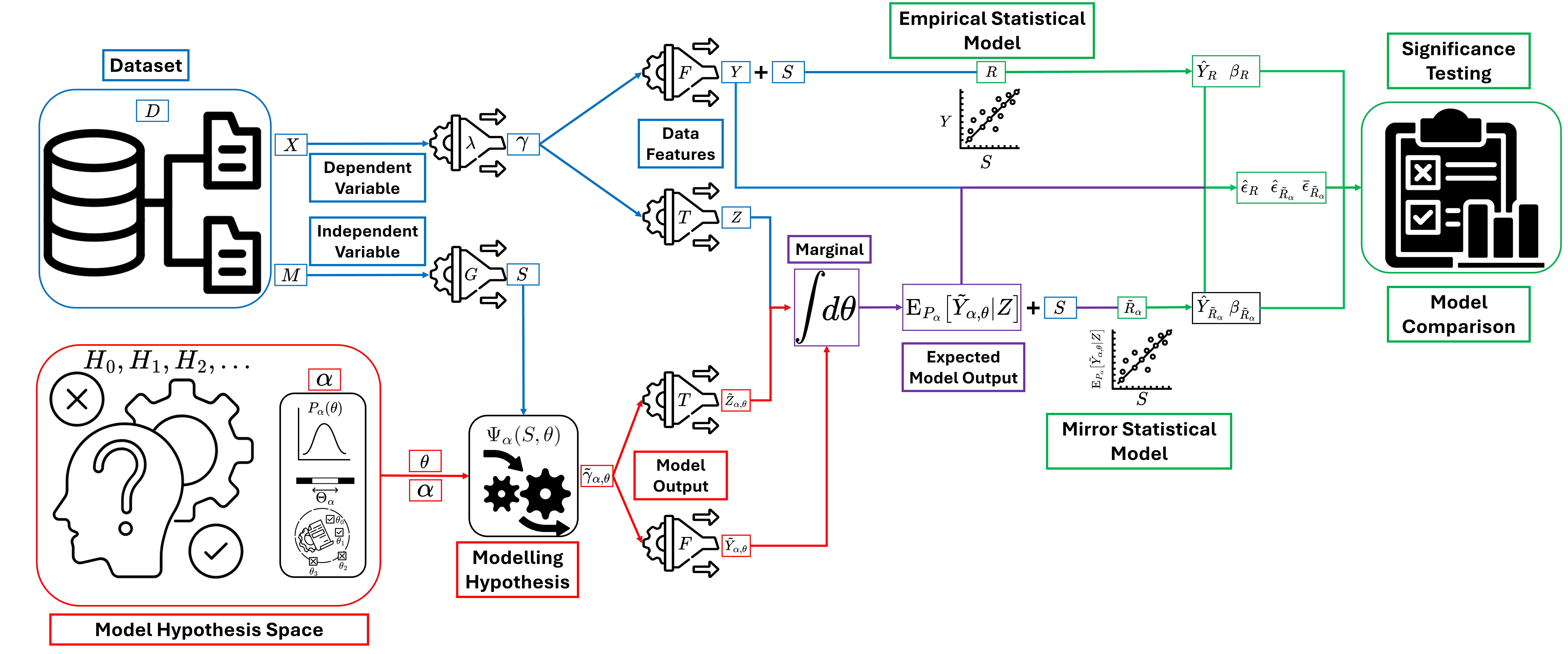}
		\caption{
			\textbf{Overview of the proposed hypothesis-first framework.} 
			Overview of the hypothesis-first framework for mechanistic model evaluation. Empirical data are first transformed into features of interest (blue), defining an empirical statistical relationship between an independent feature $S$ and a target feature $Y$. Candidate mechanistic hypotheses are then formalized as mathematical models that generate simulated data under different parameterizations (red). For each hypothesis, an expectation-based procedure computes the expected model output in $Y$ by weighting parameter configurations according to their ability to reproduce a broader or distinct feature $Z$, yielding a mirror dataset (purple). This mirror dataset is used to derive a mirror statistical model, which is directly compared to the empirical statistical model using predefined statistical criteria (green). This comparison enables systematic rejection of ill-posed, under-parameterized,  over-parameterized, or structurally invalid mechanistic hypotheses prior to parameter estimation or inference.
		}
		\label{fig:fig1}
	\end{figure}
	\botfigrule
	\subsection{Experimental Design}
	We validated our hypothesis-first framework using synthetic datasets generated from a biophysically grounded model of cortical dynamics. The aim of these experiments was to test whether the framework can (1) reject under- and over-parameterized hypotheses, (2) penalize invalid model structures, and (3) retain hypotheses that accurately capture the generative process. We designed two complementary experiments that differed in the overlap between the target feature $Y$ and the generalization feature $Z$ along with model complexity.
	
	\subsubsection{\textit{Synthetic Data Generation}}
	We used the classical Wilson-Cowan neural mass model \citep{wilson_mathematical_1973} to generate synthetic datasets with known ground truth. In Experiment 1, the overlap between $Y$ and $Z$ was minimal, which allowed us to evaluate the framework's ability to penalize over-parameterized models and compensation mechanisms. In Experiment 2, the models involved inter-connected nodes with potentially shared or differing parameter settings and $Y$ was explicitly derived from $Z$, which provided a test of the framework's sensitivity to inappropriate parameter structure and broad feature generalization respectively. In both experiments, we inverted the natural statistical relationship between the independent feature $S$ and the dependent feature $Y$ by manipulating model parameters (see Section~\ref{sec:inversion}). This ensured that only hypotheses including the relevant parameters could recover the relationship without having to rely on invalid compensation mechanisms using other parameters. 
	
	\subsubsection{\textit{Framework Implementation}}
	Because the synthetic generative relationship was linear, we modeled both empirical and mirror relationships using ordinary least squares (OLS) regression. The empirical model was defined as:
	\begin{equation}
		Y = \hat{\beta}_0 + \hat{\beta}_1 S + \hat{\epsilon}_R
	\end{equation}
	\noindent where $\hat{\beta}_0$ and $\hat{\beta}_1$ are regression coefficients and $\hat{\epsilon}_R$ denotes residuals. Mirror models $\tilde{R}_\alpha$ were fitted using expected simulated outputs $\tilde{Y}_\alpha$ derived under feature-generalization constraints.
	
	For numerical integration over the parameter space $\Theta_\alpha$, we implemented a quasi-Monte Carlo algorithm that approximates the integrals as sums over a fixed number of low-discrepancy quasi-random samples, drawn using Scipy's engine for generating Sobol sequences (QMC module), that maximize coverage of the parameter space for a given fixed computational budget \citep{SOBOL196786,leobacher_introduction_2014,virtanen_scipy_2020}. This procedure yielded a numerical approximation of the expected model output:
	\begin{equation}
		\tilde{Y}_\alpha = \int_{\theta \in \Theta_\alpha} \tilde{Y}_{\alpha,\theta} \, P_\alpha(\theta \mid Z) \, d\theta
	\end{equation}
	\noindent where $P_\alpha(\theta \mid Z)$ denotes the parameter likelihood of each point in the parameter space defined by how well its associated $\tilde{Z}_{\alpha,\theta}$ output matches the empirical feature $Z$.
	
	Empirical and mirror models were then compared with respect to their regression coefficients $(\beta_0, \beta_1)$, residual distribution, residual mean, and residual variance  (residuals comparison being for $\hat{\epsilon}_R$ vs. $\bar{\epsilon}_{R_{\alpha}}$). Differences in residual distributions were evaluated using the Baumgartner-Weiss-Schindler (BWS) test \citep{neuhauser_exact_2005}. Differences in residual means were assessed with a paired sample $t$-test, and differences in residual variances were tested using Levene's test \citep{virtanen_scipy_2020,brown_robust_1974}. Coefficients were compared based on the overlap of their confidence intervals. The significance threshold (p-value) for all tests was set to $0.01$. All statistical models were implemented using the \textit{statsmodels} package \citep{seabold2010statsmodels}.
	
	\subsubsection{\textit{The Wilson-Cowan Model}}
	
	We modeled a local cortical circuit as two interacting neuronal populations, one excitatory ($E$) and one inhibitory ($I$), described by the Wilson-Cowan differential equations:
	\begin{equation}
		\tau_e \frac{dE}{dt} = -E + (k_e - r_e E) \, f_e \left( c_{ee} E - c_{ei} I + N_{\text{input}}(t) - b_e \right)
	\end{equation}
	\begin{equation}
		\tau_i \frac{dI}{dt} = -I + (k_i - r_i I) \, f_i \left( c_{ie} E - c_{ii} I - b_i \right)
	\end{equation}
	In these equations, $c_{xy}$ are coupling coefficients ($x, y \in \{e, i\}$), $\tau_x$ are population time constants, $b_x$ are activation thresholds, and $k_x$ and $r_x$ are gain and refractory parameters. The input term $N_{\text{input}}(t)$ was defined differently depending on the experimental setup. In the network version of the model (Experiment~2), it incorporated delayed coupling between nodes in a network $\nu$:
	\begin{equation}
		N^{\text{input}}_n(t) = g \sum_{m \in \nu} A_{nm} E_m \left( t - \frac{L_{nm}}{K} \right)
	\end{equation}
	Here, $A$ is the structural connectivity matrix, $L$ is the tract-length matrix, $g$ is a global scaling factor, and $K$ is the conduction velocity. The sigmoidal activation function was defined as:
	\begin{equation}
		f_x(\eta) = \frac{1}{1 + e^{-a_x \eta}}
	\end{equation}
	Synthetic time series for $E$ were integrated using the stochastic Heun scheme, and power spectra $\gamma$ were estimated with Welch's method \citep{welch_use_1967}.
	
	\subsubsection{\textit{Parameter Bias and Relationship Inversion} \label{sec:inversion}}
	To rigorously test the framework, we introduced two controlled manipulations. First, the relationship between $S$ and $Y$ was inverted by increasing the excitatory threshold $b_e$ as a function of $S$:
	\begin{equation}
		b_e = b_e^{\text{template}} + \alpha S
	\end{equation}
	\noindent where $\alpha$ was chosen such that the sign of the $S \rightarrow Y$ relationship was reversed. Second, a constant offset was introduced in the excitatory-to-inhibitory coupling parameter $c_{ie}$:
	\begin{equation}
		c_{ie} = c_{ie}^{\text{template}} + \delta
	\end{equation}
	Both $b_e$ and $c_{ie}$ were perturbed with Gaussian noise to emulate variability in empirical data. Only hypotheses treating these parameters as free could recover the correct generative process.

	\section{Results}
	\subsection{Experiment 1: Penalizing Over-parameterization under Minimal Feature Overlap}
	
	\subsubsection{\textit{Objective}}
	The first experiment was designed to evaluate the framework's capacity to reject mechanistic hypotheses that include unnecessary free parameters when the target feature $Y$ and the generalization feature $Z$ exhibit minimal overlap. This scenario simulates a common situation in neuroimaging, where models may be over-parameterized relative to the empirical data and fit specific features in multiple ways, some that potentially don't generalize to other features of the data, through exploitation of parameter flexibility. An effective framework should penalize such hypotheses and reject them even if they reproduce $Y$ at various optimal parameter sets.
	
	\subsubsection{\textit{Design and Ground Truth}}
\begin{figure}[h!]
		\centering
		\includegraphics[width = 0.95\linewidth]{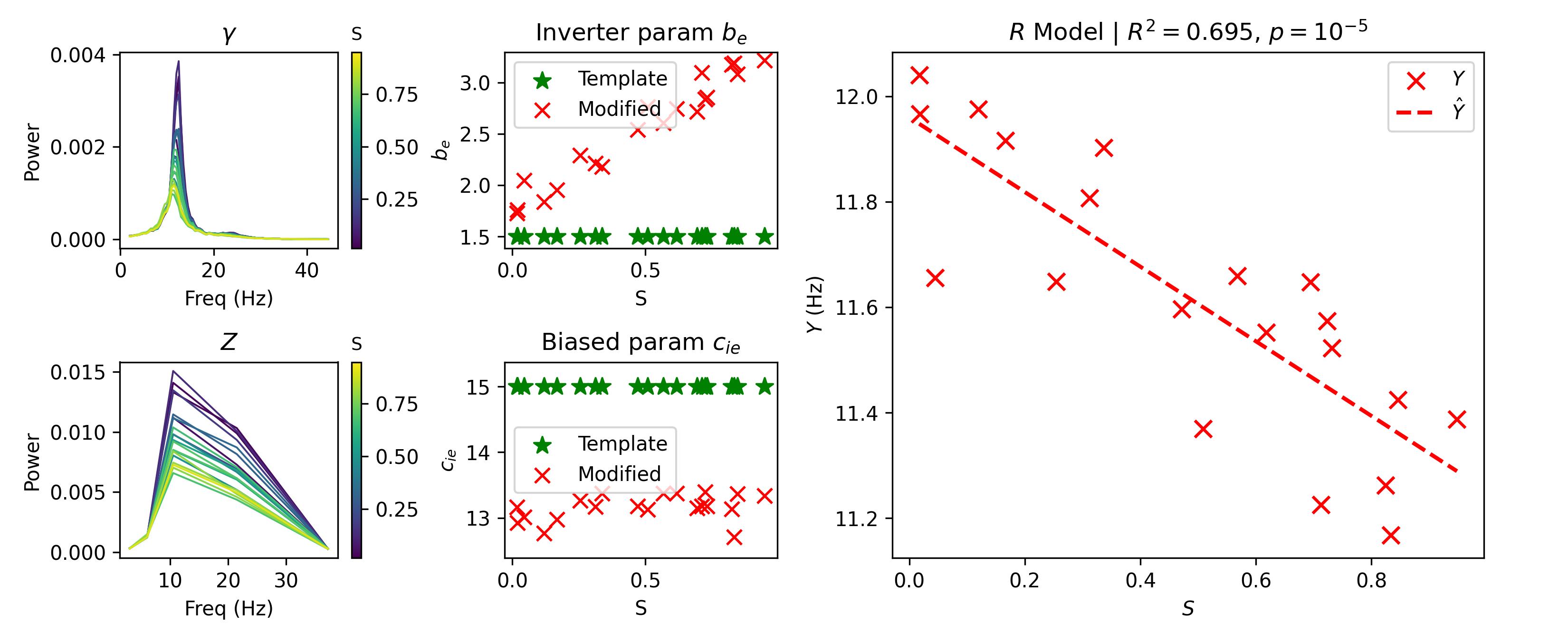}
		\caption{\textbf{Synthetic dataset summary for Experiment 1.} 
			Each column illustrates a key component of the simulation. The first column shows the simulated model output $\gamma$ and the corresponding generalization feature $Z$ across 20 data points colored according to their respective $S$ value. The middle column shows the template (green) versus modified (red) values of the inversion parameter $b_e$ (top) and bias parameter $c_{ie}$ (bottom) for each data point. The last column displays the relationship between the structural feature $S$ and the target feature $Y$, along with the fitted statistical model $R$, which corresponds to a univariate ordinary least squares linear regression. These panels illustrate all relevant components of the synthetic data with manually inverted empirical relationship between $S$ and $Y$, providing a ground truth against which mechanistic hypotheses can be evaluated.}
		\label{fig:exp1_design}
		\figdata{Data used to generate the dataset summary figure for Experiment 1.}\label{figdata:fig2}
		\figsupp[Experiment 1 dataset prior to inversion]
		{Each column illustrates a key component of the simulation. The first column shows the simulated model output $\gamma$ and the corresponding generalization feature $Z$ across 20 data points colored according to their respective $S$ value. The middle column shows the template (green) versus modified (red) values of the inversion parameter $b_e$ (top) and bias parameter $c_{ie}$ (bottom) for each data point. The last column displays the relationship between the structural feature $S$ and the target feature $Y$, along with the fitted statistical model $R$, which corresponds to a univariate ordinary least squares linear regression. These panels illustrate all relevant components of the synthetic data before manually inverting empirical relationship between $S$ and $Y$, showing the natural trend of the relationship when the inversion parameter is not adjusted to reverse it.}
		{\includegraphics[width = 0.95\linewidth]{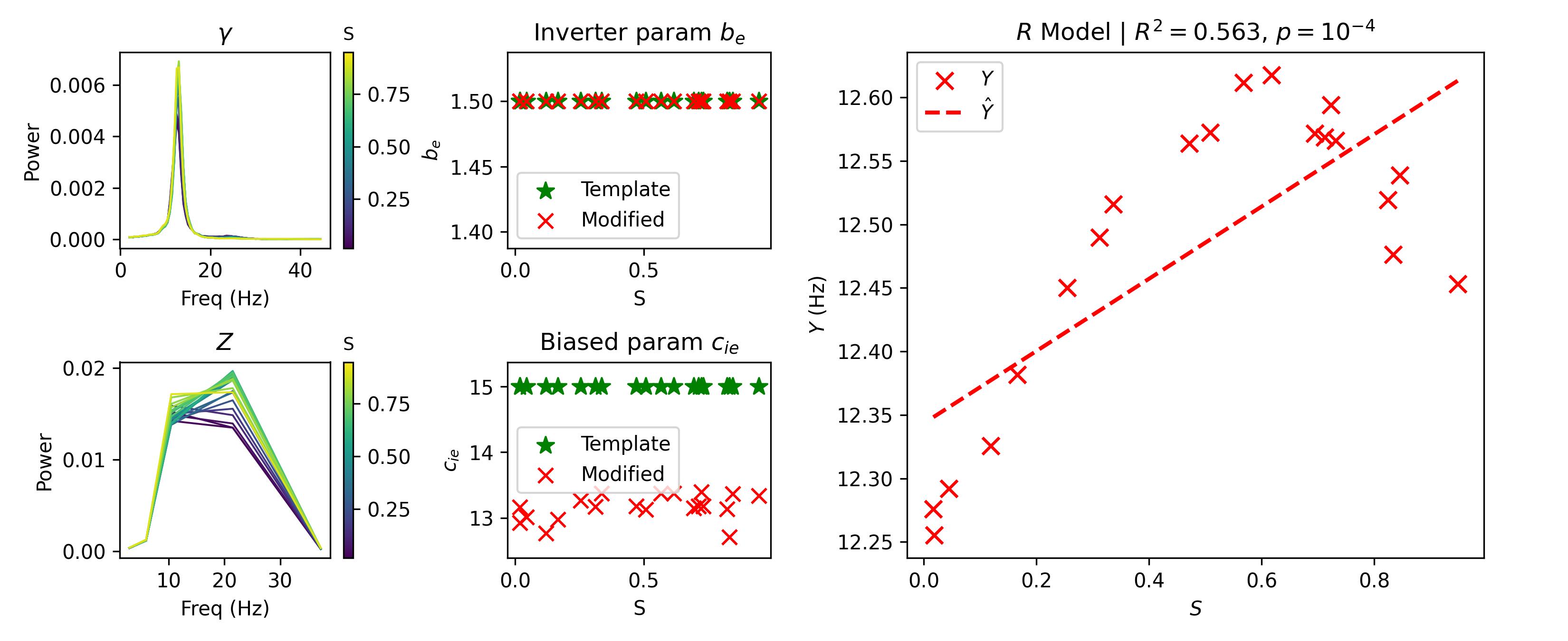}}
		\label{figsupp:exp1_design_supp}
	\end{figure}

	Synthetic datasets were generated using the Wilson-Cowan neural mass model described in the above Methods Section, with parameter manipulations designed to invert the empirical relationship between structure $S$ and function $Y$. The ground-truth generative model linked changes in the excitatory threshold $b_e$ to the structural feature $S$, such that alpha-band power decreased with increasing $S$. Formally, the manipulation was implemented as:
	\begin{equation}
		b_e = b_e^{\text{template}} + c S
	\end{equation}
	\noindent where $c < 0$ was chosen so that $\frac{dY}{dS} < 0$. Under this setup, valid mechanistic hypotheses required inclusion of $b_e$ as a free parameter to recover the true slope of the relationship without resorting to compensation. 
	The overall design of the experiment and structure of the parameter manipulations are illustrated in Figure~\ref{fig:exp1_design}. To specify minimal feature overlap, the generalization feature $Z$ was defined as spectral band power in the five canonical frequency bands together spanning the $2$-$45$~Hz range, while the target feature $Y$ corresponded to the "center of mass" frequency within a slightly broader alpha-band ($8$-$15$~Hz) than the one defined for $Z$ ($8$-$12$~Hz). This ensured that the model could not trivially reproduce $Y$ by optimizing $Z$, since the variability range of $Y$ is mostly collapsed within $Z$, and forced the evaluation to focus on generalization behavior.
	
	\subsubsection{\textit{Hypotheses}}
	The framework was applied to three mechanistic hypotheses: (i) a correctly specified model including $b_e$ and $c_{ie}$ as a free parameters, (ii) an under-parameterized model in which $b_e$ was fixed and $c_{ie}$ was free, and (iii) an over-parameterized model including additional free parameters unrelated to the ground truth ($\tau_i$ and $a_e$). For each hypothesis, expected outputs $\mathrm{E}_{P_{\alpha}}[\tilde{Y}_{\alpha, \theta}| Z]$ were computed under the feature-generalization constraint, and mirror statistical models $\tilde{R}_\alpha$ were derived as described in the above Methods Section. 
	\subsubsection{\textit{Outcomes}}

\begin{figure}[h!]
		\centering
		\includegraphics[width = 0.95\linewidth]{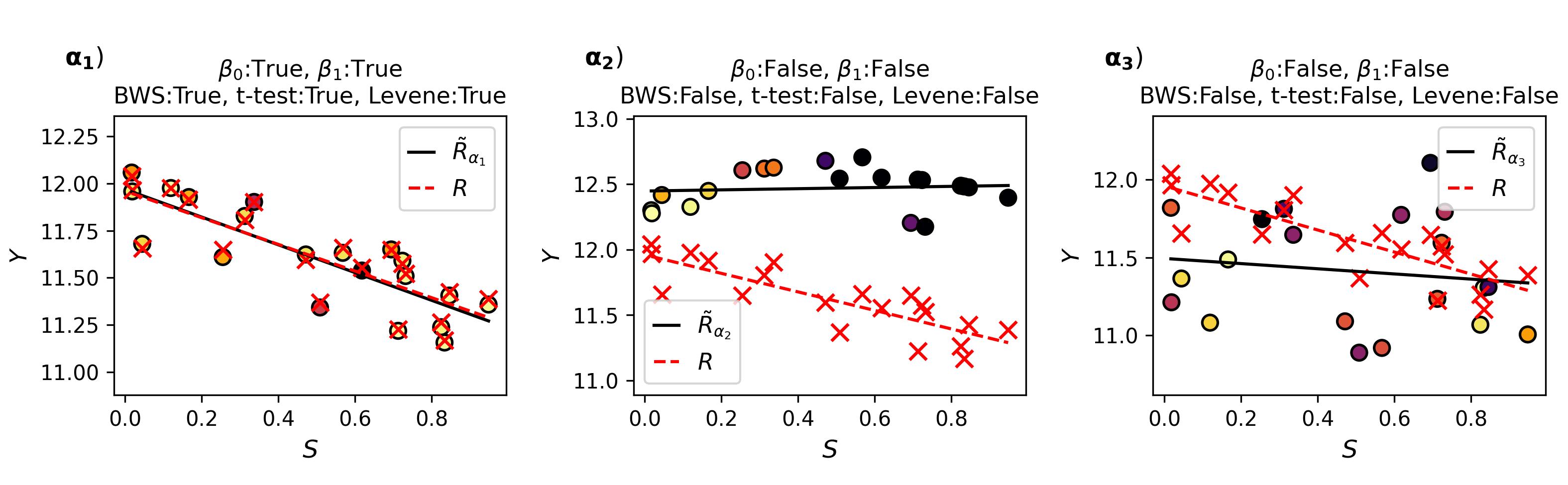}
		\caption{\textbf{Empirical and mirror statistical models for Experiment 1 hypotheses.} 
			Each panel compares the empirical statistical model $R$ (dashed red line) to the mirror statistical model $\tilde{R}_\alpha$ (full black line) generated by the framework for three mechanistic hypotheses ($\alpha_1$ to $\alpha_3$). Their associated datasets are shown as red cross and colored dots respectively. For mirror data points, the color scheme of the dots corresponds to the likelihood value from the expected model output $\tilde{Z}_{\alpha}$ ($\log_{10}$ scale), with yellow dots corresponding to high likelihood and black dots corresponding to low likelihood, see supplementary figure for examples. Binary outcomes indicate whether model coefficients (intercept, slope) and residual distributions (CDF, mean, variance) differed significantly between empirical and mirror models (details of the statistics in the source file). Hypotheses lacking key parameters ($\alpha_2$) or containing unnecessary degrees of freedom ($\alpha_3$) are rejected according to all five tests, while the true hypothesis ($\alpha_1$) is retained.}
		\label{fig:exp1_results}
		\figdata{Data used to generate the results figures for Experiment 1.}\label{figdata:fig34}
		\figsupp[Experiment 1 examples of expected $\tilde{Z}_{\alpha}$ outputs]
		{The panels show the true $Z$ feature (red) and the expected $\tilde{Z}_{\alpha}$ model output (black) for three representative data points with the highest (top), median (middle), and lowest (bottom) likelihood values. Each column corresponds to one of the three mechanistic hypotheses tested.}
		{\includegraphics[width = 0.95\linewidth]{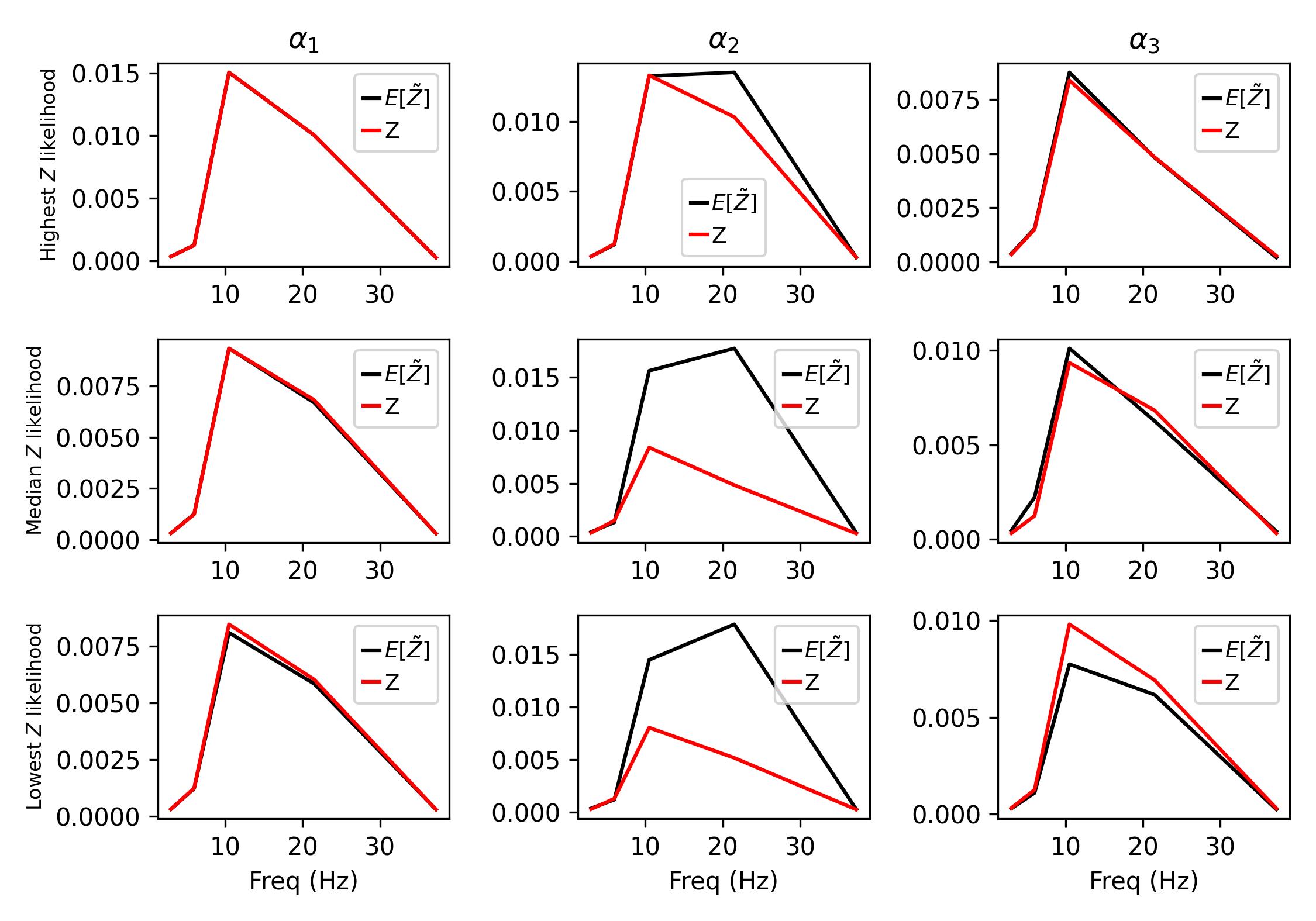}}
		\label{figsupp:exp1_results_supp}
	\end{figure}

	The main results are shown in Figure~\ref{fig:exp1_results}. The correctly specified modeling hypothesis produced mirror regression coefficients $\tilde{\beta}_0$ and $\tilde{\beta}_1$ that were statistically indistinguishable from the empirical coefficient:
	\begin{equation}
		\tilde{\beta}_0 \approx \hat{\beta}_0, \quad \tilde{\beta}_1 \approx \hat{\beta}_1 \qquad p > 0.01
	\end{equation}
	\noindent and residual distributions were indistinguishable across all tests (Figure~\ref{fig:exp1_results}, panel $\alpha_1$). \\
	
	By contrast, the under-parameterized modeling hypothesis failed to capture the inverted relationship, yielding a mirror slope (and intercept) significantly different from the empirical one in addition to residuals with non-overlapping distributions (Figure~\ref{fig:exp1_results}, panel $\alpha_2$). The over-parameterized modeling hypothesis produced mirror statistical coefficients closer to empirical ones but remained significantly different for all coefficient and residuals tests:	
	\begin{equation}
		\begin{alignedat}{2}
			&\tilde{\beta}_0 \not\approx \hat{\beta}_0, \quad \tilde{\beta}_1 \not\approx \hat{\beta}_1  \qquad&&p < 0.01\\
			&\bar{\epsilon}_{R_{\alpha}} \not\approx \hat{\epsilon}_R &&p < 0.01
		\end{alignedat}
	\end{equation}
	\noindent indicating overfitting of $Z$ at the cost of generalization performance to $Y$ (Figure~\ref{fig:exp1_results}, panel $\alpha_3$).
	
	\begin{figure}[h!]
		\centering
		\includegraphics[width = 0.8\linewidth]{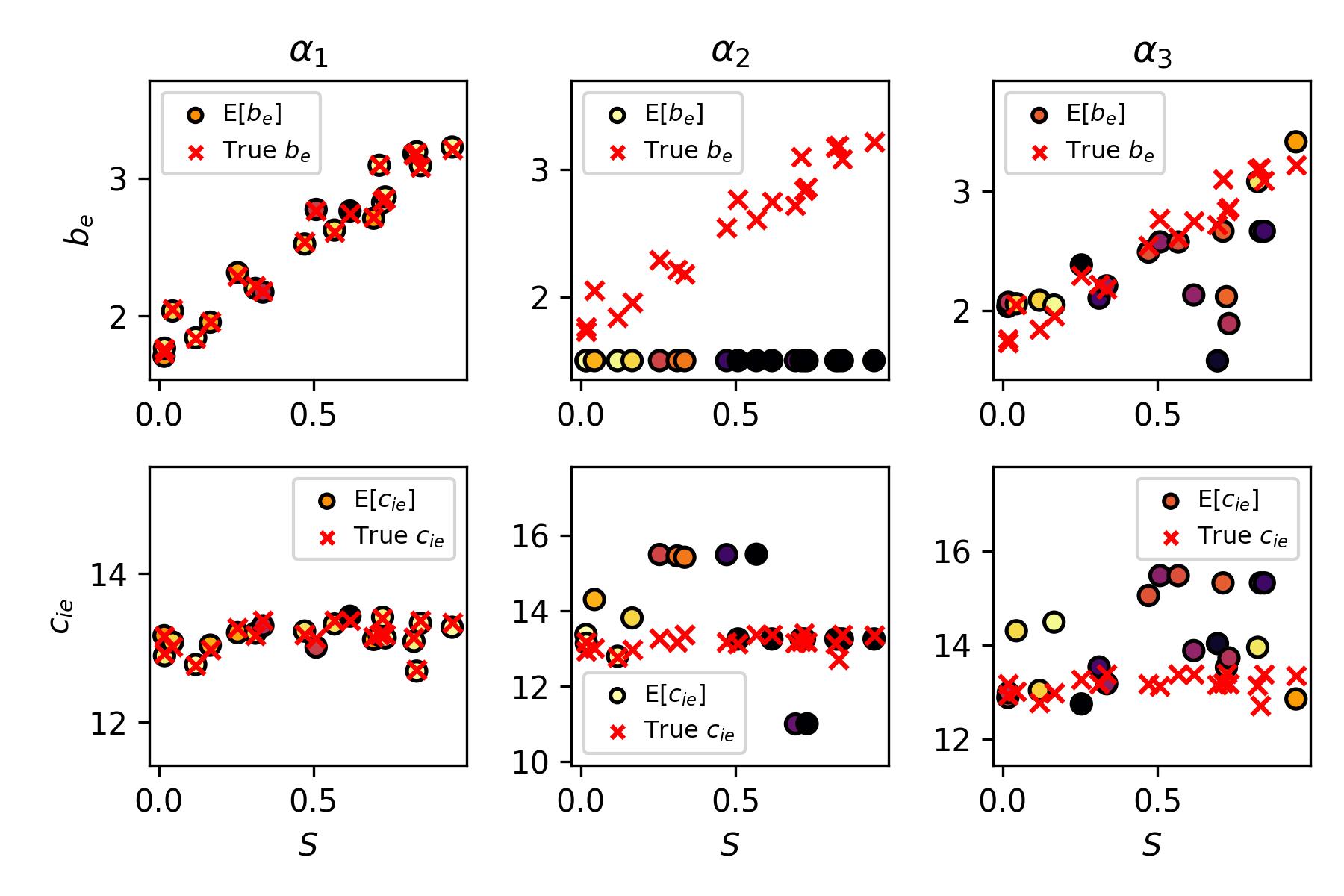}
		\caption{\textbf{Expected Parameters for Experiment 1.} 
			True parameter values (red crosses) and expected parameter estimates (colored dots; same color scheme as in Figure 3) are shown for all three hypotheses after passing through the framework. Note: These parameter values do not represent the parameter configuration associated with the $\tilde{Y}_{\alpha}$ and $\tilde{Z}_{\alpha}$ outputs shown in Figure 3 and the associated supplementary figure, instead they locate the center of mass of the likelihood distribution used to compute them within the parameter space. Overlapping dots reflect that the correct regions of the parameter space contributed most to the computation of the mirror dataset, as is the case for the true hypothesis $\alpha_1$ (left column). Mismatching dots reflects either the absence of a key free parameter, as is the case for $\alpha_2$ (middle column), or the presence of interference due to multiple solutions for reproducing $Z$, as is the case for $\alpha_3$ (right column).}
		\label{fig:exp1_parameters}
	\end{figure}
	\subsubsection{\textit{Interpretation}}	
	These results demonstrate that the framework robustly penalizes over-parameterized models even when they are equipped to successfully reproduce the target feature at an optimal parameter set. By leveraging expected model behavior under feature generalization constraints, the approach penalizes unnecessary free parameters with non-generalizing solution spaces and flags such ill-posed hypotheses for rejection. This outcome illustrates one of the framework's key advantages: it acts as a pre-inference filter that prevents the propagation of inappropriately specified models into downstream analyses.
	
	\subsection{Experiment 2: Sensitivity to Parameter Structure and Overlapping Features}
	
	\subsubsection{\textit{Objective}}
	The second experiment was designed to evaluate the framework's performance in a more complex model and when the target feature $Y$ and the generalization feature $Z$ contain overlapping information. This setting tests whether the approach can still reject inappropriate mechanistic hypotheses when the feature overlap makes flagging overfitting more challenging, and whether it remains sensitive to structural aspects of the parameter space such as invalid parameter sharing and associated compensation mechanisms.
	
	\subsubsection{\textit{Design and Ground Truth}}
\begin{figure}[h!]
		\centering
		\includegraphics[width = 0.95\linewidth]{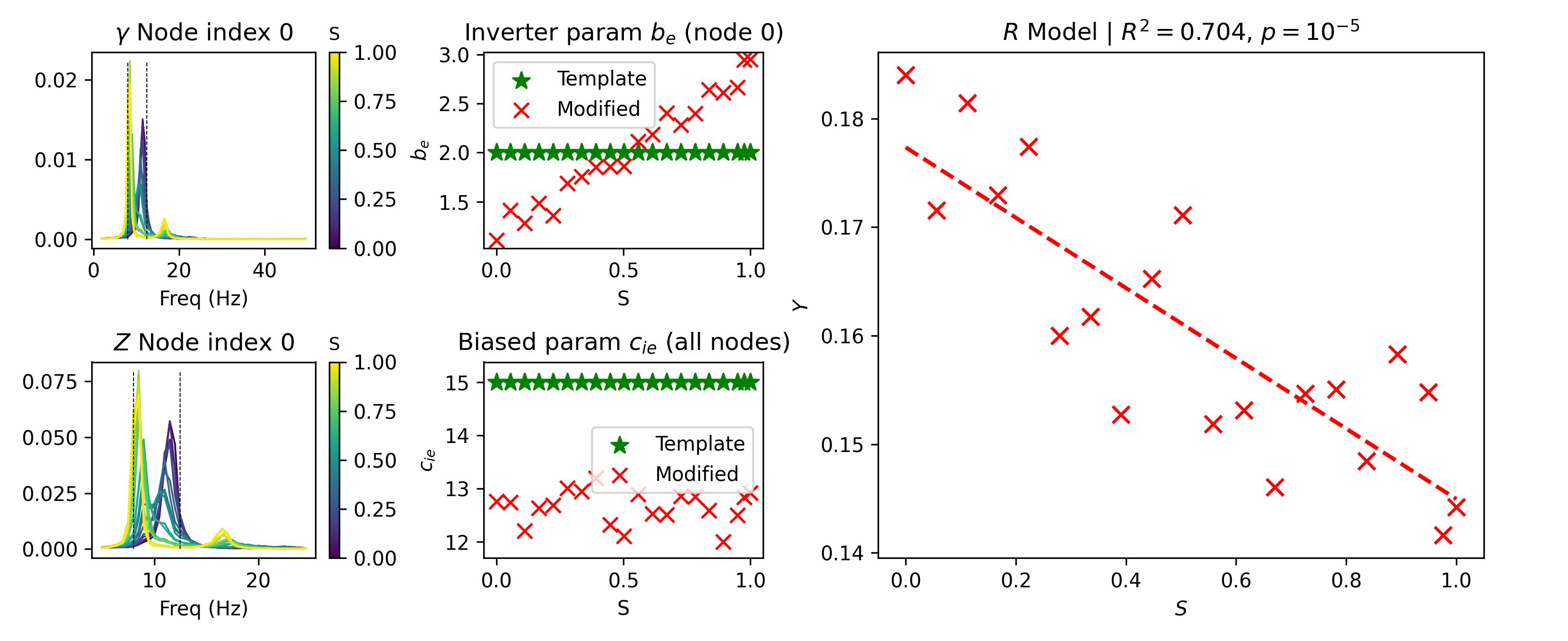}
		\caption{\textbf{Synthetic dataset summary for Experiment 2.} 
			Each column illustrates a key component of the simulation. The first column shows the simulated model output $\gamma$ and the corresponding generalization feature $Z$ (only for node $n_0$) across 20 data points colored according to their respective $S$ value. The dashed vertical lines in this first column represent the bounds of the frequency range used to compute $Y$. The middle column shows the template (green) versus modified (red) values of the inversion parameter $b_e$ (top) and bias parameter $c_{ie}$ (bottom) for each data point. The last column displays the relationship between the structural feature $S$ and the target feature $Y$, along with the fitted statistical model $R$, which corresponds to a univariate ordinary least squares linear regression. These panels illustrate all relevant components of the synthetic data with manually inverted empirical relationship between $S$ and $Y$, providing a ground truth against which mechanistic hypotheses can be evaluated. This setup introduces stronger overlap between $Y$ and $Z$ and parameter sharing constraints, enabling evaluation of the framework under more complex modeling conditions.}   
		\label{fig:exp2_design}
		\figdata{Data used to generate the dataset summary figure for Experiment 2.}\label{figdata:fig5}
		\figsupp[Experiment 2 dataset prior to inversion.]
		{\textbf{Synthetic dataset summary before inversion for Experiment 2.} 
			Each column illustrates a key component of the simulation. The first column shows the simulated model output $\gamma$ and the corresponding generalization feature $Z$ (only for node $n_0$) across 20 data points colored according to their respective $S$ value. The dashed vertical lines in this first column represent the bounds of the frequency range used to compute $Y$. The middle column shows the template (green) versus modified (red) values of the inversion parameter $b_e$ (top) and bias parameter $c_{ie}$ (bottom) for each data point. The last column displays the relationship between the structural feature $S$ and the target feature $Y$, along with the fitted statistical model $R$, which corresponds to a univariate ordinary least squares linear regression. These panels illustrate all relevant components of the synthetic data before manually inverting empirical relationship between $S$ and $Y$, showing the natural trend of the relationship when the inversion parameter is not adjusted to reverse it.}
		{\includegraphics[width = 0.95\linewidth]{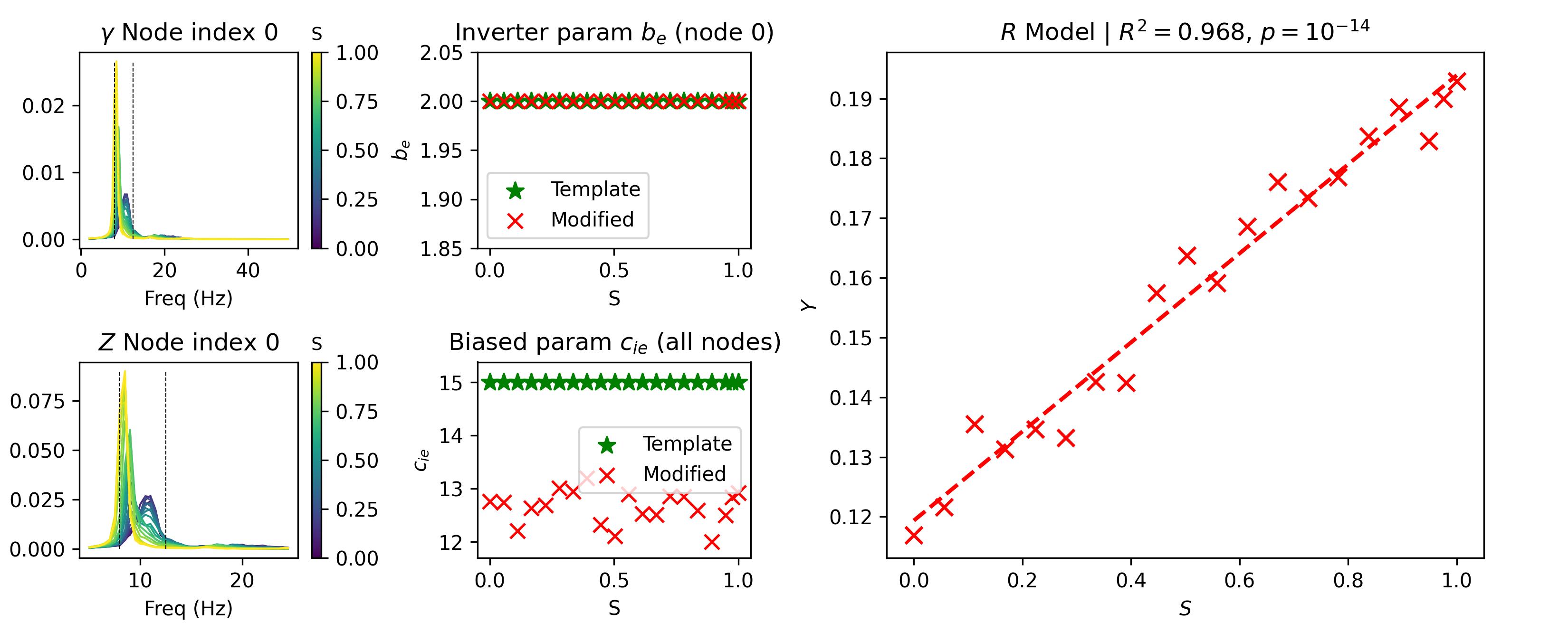}}
		\label{figsupp:exp2_design_supp}
		\figsupp[Connectivity matrices used in Experiment 2.]
		{Weight ($A$) and tract-length ($L$) matrices are shown for three representative datasets corresponding to the minimum (left), median (middle), and maximum (right) values of the structural feature $S$. The top row displays the weight matrices and the bottom row the tract-length matrices. These illustrate how the connection strengths of node $n_0$ were systematically scaled across datasets, creating a controlled linear variation in $S$ while maintaining the same base topology and distance structure.}
		{\includegraphics[width = 0.95\linewidth]{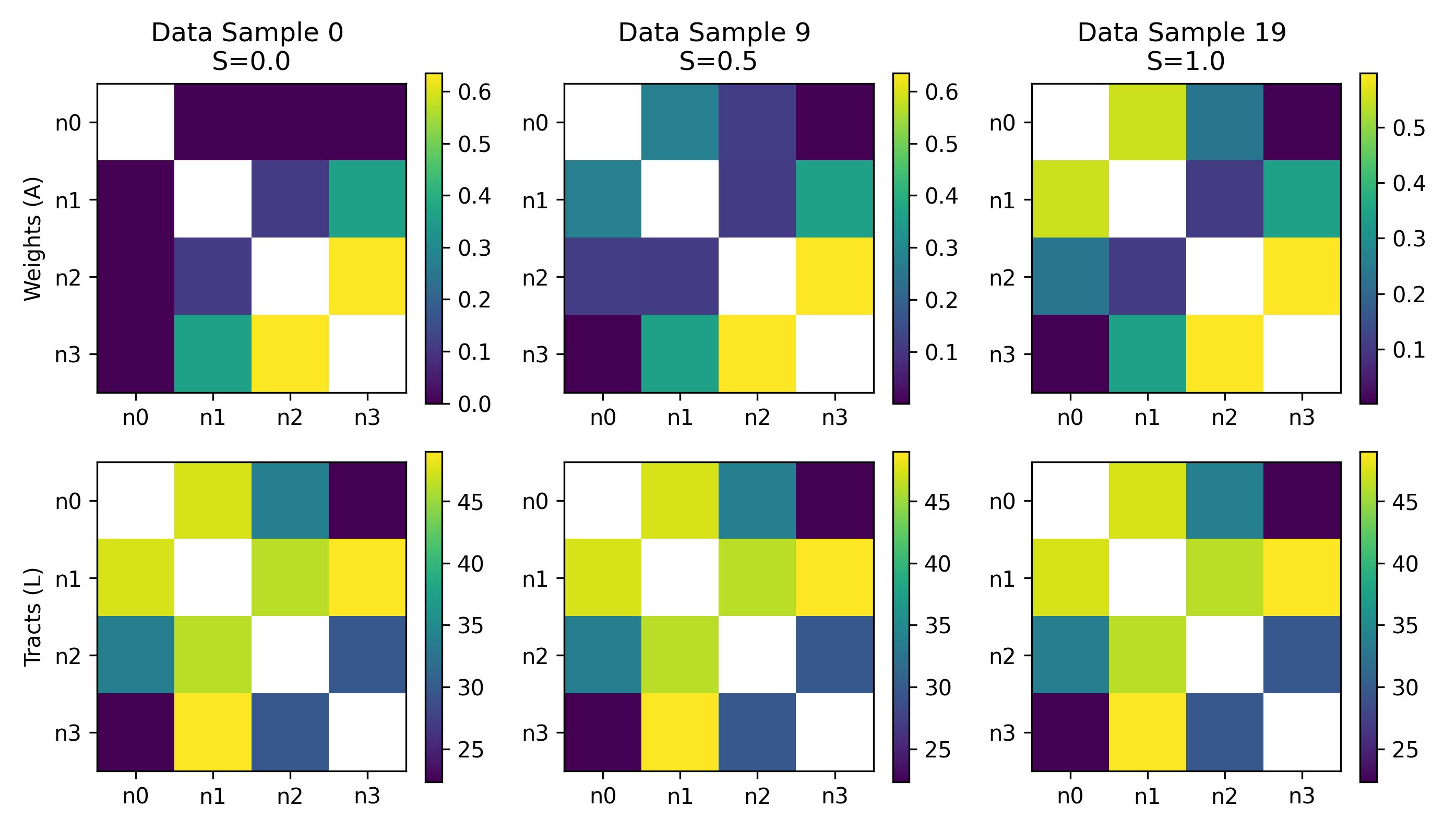}}
		\label{figsupp:connectivity}
	\end{figure}

	We modeled a coupled system of four interconnected nodes with connectivity matrices $M = [A, L]$, where $A$ represents coupling strengths and $L$ the tract lengths. The structural feature $S$ was defined as the degree of the coupling strength matrix for the first node $n_0$:
	\begin{equation}
		S = \sum_{n=0}^{3} A_{0n}
	\end{equation}
	The generalization feature $Z$ was chosen as a truncated, normalized power spectrum. Each node's power was first normalized by the total power across all nodes and then truncated to the $5$–$25$\,Hz band, as shown in Figure~\ref{fig:exp2_design} (bottom left panel). The target feature $Y$ was then defined as the sum of this same normalized power within the $8$–$13$\,Hz band for node $n_0$:
	\begin{equation}
		Y = \sum_{\omega \in \Omega} Z_{0,\omega}, \quad \Omega = \{8, 8.5, \dots, 12.5\}\,\text{Hz}
	\end{equation}
	Dataset generation followed the same procedure as Experiment~1, except that connectivity weights and distances were drawn from distributions generally close to those of empirical human connectomes. A linear relationship between $S$ and $Y$ was again imposed by scaling the sum of inputs to the first node. As before, this relationship was inverted by modulating the excitatory threshold parameter $b_e$ as a function of $S$, but in this case, the inversion was applied only to node $n_0$, while all other nodes retaining the template parameter value of $b_e$ (Figure~\ref{fig:exp2_design}, middle column). The change in the bias parameter $c_{ie}$ was performed in an "all nodes shared" parameter structure.
	
	\subsubsection{\textit{Hypotheses}}
	Five mechanistic hypotheses were tested, differing in parameter inclusion, sharing structure, and model degrees of freedom. Hypotheses $\alpha_1$ and $\alpha_2$ mirrored those of Experiment~1, while $\alpha_3$–$\alpha_5$ introduced new forms of over-parameterization and/or mis-specification. For example, $\alpha_3$ included the correct parameters but without sharing across nodes, resulting in an over-complete model with unnecessary degrees of freedom. The hypothesis $\alpha_4$ included the same unnecessary free parameters as $\alpha_3$ in Experiment~1, but fully defined in an invalid "all nodes shared" parameter structure. The hypothesis $\alpha_5$ mirrors $\alpha_1$ by containing only the correct free parameters, but it was defined with an invalid simplifying assumption that forced all nodes to share the same $b_e$, thereby introducing a constraint results in invalid spectral patterns in nodes other than $n_0$, which is information present in $Z$, despite being fully equipped for reproducing $Y$ on its own.
	
	\subsubsection{\textit{Outcomes}}
\begin{figure}[h!]
		\centering
		\includegraphics[width = 0.95\linewidth]{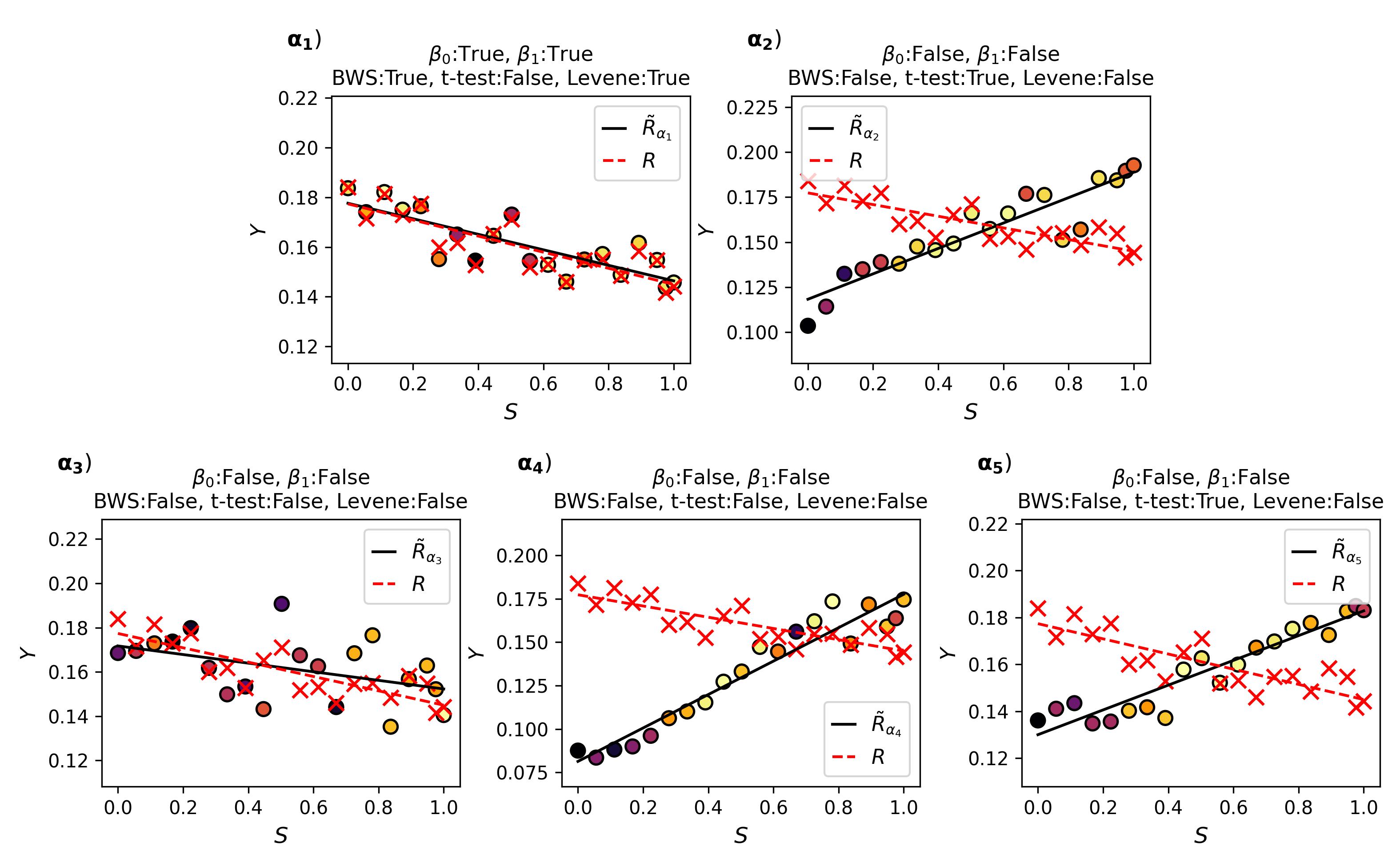}
		\caption{\textbf{Empirical and mirror statistical models for Experiment 2 hypotheses.} 
			Each panel compares the empirical statistical model $R$ (dashed red line) to the mirror statistical model $\tilde{R}_\alpha$ (full black line) generated by the framework for five mechanistic hypotheses ($\alpha_1$ to $\alpha_5$). Their associated datasets are shown as red cross and colored dots respectively. For mirror data points, the color scheme of the dots corresponds to the likelihood value from the expected model output $\tilde{Z}_{\alpha}$ ($\log_{10}$ scale), with yellow dots corresponding to high likelihood and black dots corresponding to low likelihood, see supplementary figure for examples. Binary outcomes indicate whether model coefficients (intercept, slope) and residual distributions (CDF, mean, variance) differed significantly between empirical and mirror models (details of the statistics in the source file). Despite the stronger feature overlap, the framework correctly rejects under-parameterized, over-parameterized, or structurally invalid models, and retains only the true mechanistic hypothesis ($\alpha_1$).}
		\label{fig:exp2_results}
		\figdata{Data used to generate the results figures for Experiment 2}\label{figdata:fig67}
		\figsupp[Experiment 2 examples of expected Z outputs]
		{The panels show the true $Z$ feature (red) and the expected $\tilde{Z}_{\alpha}$ model output (black) for three representative data points with the highest (top), median (middle), and lowest (bottom) likelihood values. Only the values for $n_0$ are shown for visual clarity. Each column corresponds to one of the five mechanistic hypotheses tested.}
		{\includegraphics[width = 0.95\linewidth]{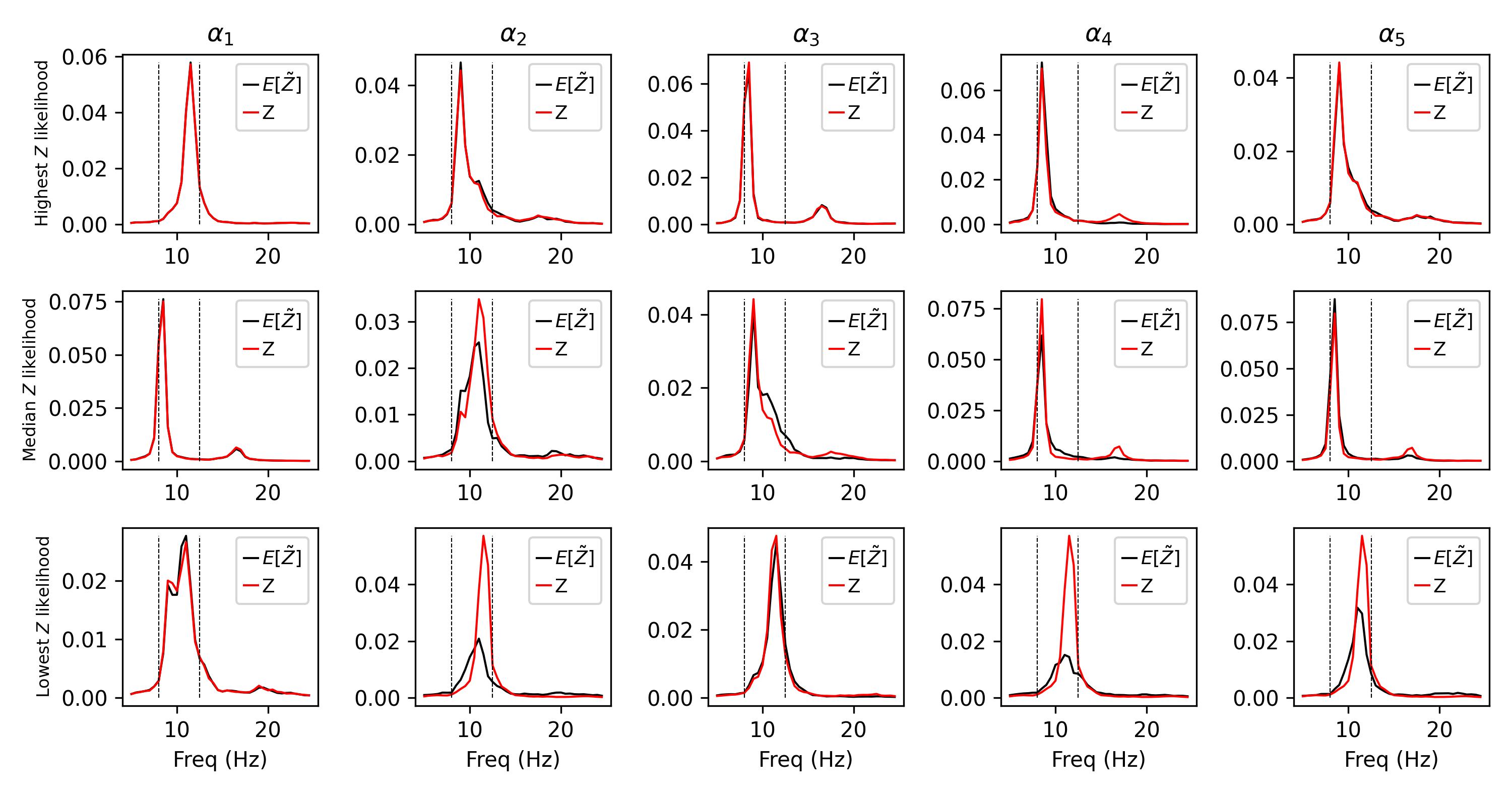}}
		\label{figsupp:exp2_results_supp}
	\end{figure}

\begin{figure}[h!]
		\centering
		\includegraphics[width = 0.95\linewidth]{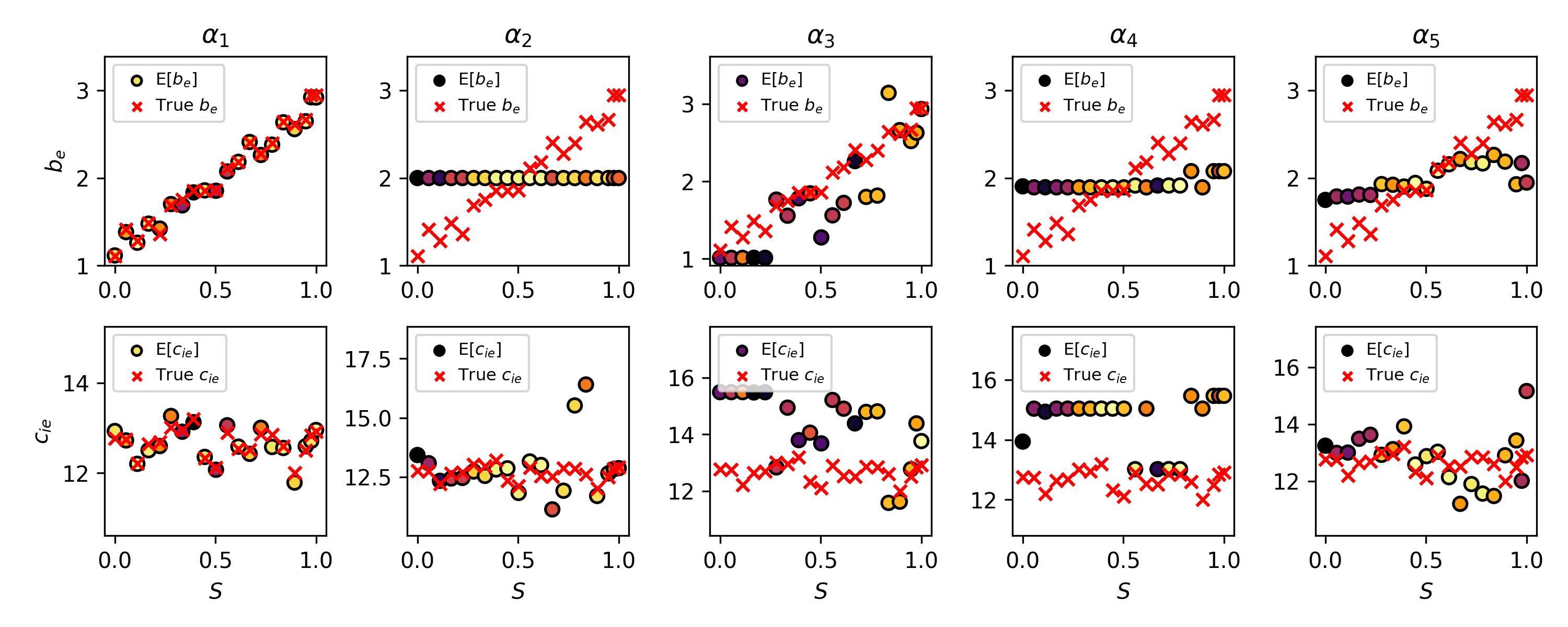}
		\caption{\textbf{Expected Parameters for Experiment 2.} 
			True parameter values (red crosses) and expected parameter estimates (colored dots; same color scheme as in Figure 6) are shown for all five hypotheses after passing through the framework. Note: These parameter values do not represent the parameter configuration associated with the $\tilde{Y}_{\alpha}$ and $\tilde{Z}_{\alpha}$ outputs shown in Figure 5 and associated supplementary figure, instead they locate the center of mass of the likelihood distribution used to compute them within the parameter space. Overlapping dots reflect that the correct regions of the parameter space contributed most to the computation of the mirror dataset, as is the case for the true hypothesis $\alpha_1$ (left column). Mismatching dots reflects either the absence of a key free parameter, as is the case for $\alpha_2$ (middle column), or the presence of interference due to multiple, sometimes conflicting, solutions for reproducing the different aspects of $Z$ across all nodes, as is the case for $\alpha_3$ through $\alpha_5$ (middle through rightmost columns).}
		\label{fig:exp2_parameters}
	\end{figure}

	Results are shown in Figures~\ref{fig:exp2_results} and~\ref{fig:exp2_parameters}. The framework successfully rejected the under-parameterized hypothesis $\alpha_2$, which lacked the inverter parameter and resulted in a mirror statistical model significantly different from the empirical model in all tests except the residuals $t$-test (Figure~\ref{fig:exp2_results}, top right panel).
	
	\textcolor{white}{-}
	
	The true hypothesis $\alpha_1$ produced mirror coefficients indistinguishable from those of the empirical model:
	\begin{equation}
		\tilde{\beta}_0 \approx \hat{\beta}_0, \quad \tilde{\beta}_1 \approx \hat{\beta}_1  \qquad p > 0.01
	\end{equation}
	\noindent and the distribution of $\bar{\epsilon}_{\tilde{R}_{\alpha_1}}$ was indistinguishable ($p>0.01$) from $\hat{\epsilon}_R$ on all tests except the $t$-test for residual means. The paired sample $t$-test is very sensitive to any systematic differences and picked up on a very small systematic overestimation from expectation approximations, which is visible in the top left panel of Figure~\ref{fig:exp2_results}: the colored dots are consistently slightly above the red crosses. By contrast, the over-parameterized $\alpha_3$ and $\alpha_4$ hypotheses exhibited significantly significantly different coefficient estimates and residual distributions:
	\begin{equation}
		\begin{alignedat}{2}
			&\tilde{\beta}_0 \not\approx \hat{\beta}_0, \quad \tilde{\beta}_1 \not\approx \hat{\beta}_1  \qquad&&p < 0.01\\
			&\bar{\epsilon}_{R_{\alpha}} \not\approx \hat{\epsilon}_R &&p < 0.01
		\end{alignedat}
	\end{equation}
	Finally, the mis-specified $\alpha_5$ hypothesis — although capable of recovering the correct relationship properties captured by $R$ — was flagged by the framework due to significant discrepancies in residual distributions and statistical model coefficients, reflecting its failure to generalize across the full spectral feature space that includes the other nodes that are not involved within $R$.
	
	\subsubsection{\textit{Interpretation}}
	These findings demonstrate that the framework remains robust even when feature overlap is high and compensation mechanisms could mask model mis-specification when looking only at $Y$ outputs. By incorporating feature generalization and mirror-model comparison, the approach correctly rejects hypotheses that are under-parameterized, over-parameterized, or structurally invalid. Importantly, it identifies cases where a model could fit the target feature $Y$ but fails to account for a broader feature set $Z$, thus revealing false-positive solutions that would likely go undetected in standard parameter estimation workflows. This highlights the framework's value as a rigorous pre-inference filter for mechanistic modeling in neuroimaging.
	
	\section{Discussion}
	The present study introduces a hypothesis-first framework for mechanistic modeling that addresses a fundamental challenge in neuroimaging: how to move from statistical description to mechanistic explanation in a way that is both rigorous and accessible. While statistical models remain indispensable for quantifying relationships between neural features, they offer limited insight into the underlying biological processes that generate those relationships. Mechanistic models, by contrast, can embody explicit hypotheses about causal mechanisms, but their adoption has been slowed by methodological complexity, the risk of model mis-specification, and the difficulty of assessing model appropriateness prior to parameter estimation/inference.
	
	Our framework directly targets these limitations by re-framing mechanistic modeling as a problem of hypothesis formulation and rejection. The key innovation is the evaluation of a model's expected behavior under a novel approach to applying feature generalization constraints, followed by the construction of mirror statistical models that allow direct comparison with empirical observations. This approach makes it possible to reject models that are under- or over-parameterized, that fail to capture relevant biological processes, or that rely on invalid simplifying assumptions, all before attempting to interpret the results of parameter estimation/inference. By shifting the focus upstream, our method functions as a pre-inference filter that ensures that subsequent inference efforts are directed only toward mechanistically plausible models.
	
	The results from our two experiments demonstrate the power and flexibility of this approach. In Experiment~1, where the overlap between the target feature $Y$ and the generalization feature $Z$ was deliberately minimized, the framework successfully penalized over-parameterized models that had all the necessary degrees of freedom to reproduce $Y$ but failed to do so while also generalizing beyond it. This finding illustrates a critical property of the method: its ability to identify unnecessary degrees of freedom that contribute to overfitting without improving explanatory adequacy given the data. In Experiment~2, where $Y$ and $Z$ were strongly related and where parameter compensation could have masked model mis-specification, the framework still correctly rejected inappropriate hypotheses, including those with structurally invalid parameter-sharing assumptions. These results demonstrate that feature generalization and mirror-model comparison remain effective even in challenging conditions where traditional approaches might yield false positives.
	
	Importantly, this framework does not replace parameter estimation or simulation-based inference but rather complements them in an accessible manner. By filtering out implausible models early, it reduces the risk of misleading inferences, improves the interpretability of fitted/inferred parameters, and provides a principled basis for subsequent modeling decisions. It also lowers the barrier to entry for researchers without extensive training in dynamical systems modeling by translating complex methodological choices into explicit, testable hypotheses, thereby leveraging their expertise to implement rigor.
	
	Several limitations and opportunities for future work should be noted. First, although our demonstrations relied on synthetic data, future work will extend the framework to empirical neuroimaging datasets, where additional challenges such as measurement noise, individual variability, and model mismatch are more pronounced. Second, while the present implementation focused on linear statistical models and a limited set of dynamical systems, the underlying principles are general and can be extended to nonlinear statistical models, higher-dimensional parameter spaces, and more complex biophysical models. Finally, future efforts will integrate this framework with automated model selection pipelines and simulation-based inference approaches, creating a unified workflow that spans hypothesis formulation, model rejection, parameter estimation, and predictive validation.
	
	In conclusion, the framework presented here offers a practical and rigorous approach to mechanistic modeling in neuroimaging. By centering the process on hypothesis formulation and rejection rather than parameter estimation/inference alone, it provides a principled bridge between statistical description and mechanistic explanation. We anticipate that this approach will not only make mechanistic modeling more accessible to the broader neuroscience community but will also help advance our understanding of the biological mechanisms underlying complex neural phenomena.

	\bibliography{references}
	
	\newpage
	\section{Supplemental}
	
	\subsection{Framework Formal Definitions \& Derivations}
	The goal of this section is to formalize the core principles of the proposed framework. Specifically, we aim to express how mechanistic hypotheses $\alpha$ are mapped into expected feature spaces, how these predictions are compared to empirical observations, and how mirror statistical models are derived to evaluate model adequacy before parameter inference.
	
	Let the dataset be defined as $D = \{ D^0, \dots, D^N \}$, consisting of $N$ samples, where each sample $D^n = (X^n, M^n)$ contains a dependent variable $X$ (e.g., neural activity) and an independent variable $M$ (e.g., structural property) that is hypothesized to influence the generation of $X$.
	
	We define a chain of feature mappings that progressively transform raw observations into features of interest:
	\begin{align*}
		\gamma = \lambda(X)
	\end{align*}
	
	\noindent where $\gamma$ is an intermediate representation of $X$ (for instance, a spectral representation or latent state) obtained by applying a feature extraction function $\lambda$. Similarly, we define a feature of the independent variable:
	\begin{align*}
		S = G(M)
	\end{align*}
	
	\noindent where $S$ is derived from $M$ via a feature function $G$ (for example, degree of a connectivity matrix). From $\gamma$, we derive two distinct features:
	\begin{align*}
		Y = F(\gamma) \quad \text{and} \quad Z = T(\gamma)
	\end{align*}
	
	\noindent where $Y$ is the empirical feature of interest (used to test the statistical model) and $Z$ is an auxiliary feature used for mechanistic modeling and generalization. 
	
	\noindent \textit{Note: The motivation for defining this feature hierarchy is to enable parallel statistical and mechanistic modeling pipelines that converge on a shared empirical target.}
	
	We assume that a statistical model $R$ has been fitted to relate $S$ and $Y$:
	\begin{align*}
		Y = \hat{Y}_R + \hat{\epsilon}_R
	\end{align*}
	
	\noindent where $\hat{Y}_R = R(S)$ is the predicted feature and $\hat{\epsilon}_R$ are residuals. Residuals for models predicting $Z$ will be denoted $\delta$.
	
	Let $\Psi_\alpha$ denote a mechanistic model defined under hypothesis $\alpha$ with parameters $\theta \in \Theta_\alpha$. The model maps $S$ to $\gamma$:
	\begin{align*}
		\tilde{\gamma}_{\alpha, \theta} = \Psi_\alpha(S, \theta)
	\end{align*}
	
	\noindent From this, we obtain the modeled feature $\tilde{Z}$:
	\begin{align*}
		\tilde{Z}^i_{\alpha, \theta} &= T\big( \tilde{\gamma}^i_{\alpha, \theta} \big)\\
		Z^i &= \tilde{Z}^i_{\alpha, \theta} + \tilde{\delta}^i_{\alpha, \theta}
	\end{align*}
	
	We define the posterior likelihood of parameters $\theta$ under hypothesis $\alpha$ for a given sample $i$ using Bayes' theorem:
	\begin{align*}
		P_\alpha(\theta | Z^i) = \frac{P_\alpha(Z^i | \theta) P_\alpha(\theta)}{P_\alpha(Z^i)}
	\end{align*}
	
	Because $\tilde{Z}$ is generated through $\tilde{\gamma}$, we can also derive $\tilde{Y}$:
	\begin{align*}
		\tilde{Y}^i_{\alpha, \theta} &= F\big( \tilde{\gamma}^i_{\alpha, \theta} \big)\\
		Y^i &= \tilde{Y}^i_{\alpha, \theta} + \tilde{\epsilon}^i_{\alpha, \theta}
	\end{align*}
	
	We define the expected residual as:
	\begin{align*}
		\tilde{\epsilon}^i_{\alpha} &= Y^i - \mathrm{E}_{P_\alpha}[\tilde{Y}^i_{\alpha, \theta} | Z^i]\\
		&= Y^i - \int_{\theta \in \Theta_\alpha} \tilde{Y}^i_{\alpha, \theta} P_\alpha(\theta | Z^i) \, d\theta
	\end{align*}
	
	From this expectation, we construct a \textit{mirror statistical model} $\tilde{R}_\alpha$ that captures the expected relationship implied by hypothesis $\alpha$:
	\begin{align*}
		\hat{Y}_{\tilde{R}_\alpha} &= \tilde{R}_\alpha(S)\\
		\tilde{Y}_\alpha &= \hat{Y}_{\tilde{R}_\alpha} + \hat{\epsilon}_{\tilde{R}_\alpha}\\
		\bar{\epsilon}_{\tilde{R}_\alpha} &= Y - \hat{Y}_{\tilde{R}_\alpha}
	\end{align*}
	
	By statistically comparing $R$ and $\tilde{R}_\alpha$ (e.g., intercept, slope, and residual properties), we assess whether hypothesis $\alpha$ can reproduce the observed statistical relationship. This enables rejection of ill-posed mechanistic models before performing parameter inference.
	
	\subsection{Challenges of Numerical Approximations}
	The integral above is often intractable because $P_\alpha(\theta | Z^i)$ rarely has a closed-form solution. It is therefore approximated using numerical methods. One approach is Monte Carlo integration over a set of $q$ parameter samples $\theta_Q = \{ \theta_0, \dots, \theta_q \} \sim \Theta_\alpha$:
	\begin{align*}
		\mathrm{E}_{P_\alpha}[\tilde{Y}^i_{\alpha, \theta} | Z^i] \approx \sum_{j=0}^q \tilde{Y}^i_{\alpha, \theta_j} \frac{P_\alpha(Z^i | \theta_j) P_\alpha(\theta_j)}{\sum_{k=0}^q P_\alpha(Z^i | \theta_k) P_\alpha(\theta_k)}
	\end{align*}
	
	This approximation introduces potential bias depending on the sampling strategy, especially in high-dimensional parameter spaces or when $q$ is small. In such cases, importance sampling (oversampling high-likelihood regions) or simulation-based inference (SBI) techniques may improve estimation, but they introduce their own challenges, such as the need to normalize posterior estimates. Further, computing this expectation requires both the parameter likelihood and the $\tilde{Y}_{\alpha, \theta} $ outputs, thus training a likelihood/posterior estimator (e.g., with SBI) only addresses half of the intractability issues.
	
	Our framework is agnostic to the specific approximation technique used. The primary goal is not to fully recover the posterior distribution but to provide a principled means to evaluate whether a mechanistic model is adequate for further inference.

	\subsection{Template Parameters \& Bounds}

	\begin{table}[h!]
		\centering
		\begin{tabular}{c c c c}
			\toprule
			Parameter & Biological Reference & Experiment 1 & Experiment 2\\ \hline
			$c_{ee}$ & Excitatory to excitatory coupling & 15 & 16\\ 
			$c_{ei}$ & Inhibitory to excitatory coupling & 12 & 12\\ 
			$c_{ie}$ & Excitatory to inhibitory coupling & 15; [11, 15.5] & 15; [11, 17]\\ 
			$c_{ii}$ & Inhibitory to inhibitory coupling & 0.5 & 0\\ 
			$\tau_e$ & Excitatory time-constant & 8 & 8\\ 
			$\tau_i$ & Inhibitory time-constant & 35; [10, 50]  & 30; [10, 50]\\ 
			$b_e$ & Excitatory activation function threshold & 1.5; [1.375, 4.5] & 2; [1, 3.5]\\ 
			$b_i$ & Inhibitory activation function threshold & 3 & 3\\ 
			$a_e$ & Excitatory activation function slope & 0.85; [0.5, 2.5] & 1.0; [0.5, 2.5]\\ 
			$a_i$ & Inhibitory activation function slope & 1.25 & 1.25\\ 
			$r_i$ & Inhibitory refractory factor & 1 & 1\\ 
			$r_e$ & Excitatory refractory factor & 1 & 1\\ 
			$k_e$ & Maximum excitatory response & 1 & 1\\ 	
			$k_i$ &	Maximum inhibitory response & 1 & 1\\ 	
			$\sigma_E$  & Noise variance of $E$ ($\log_{10}$ scale) & -4.5 & -4.5\\ 
			$\sigma_I$  & Noise variance of $I$ ($\log_{10}$ scale) & -4.5 & -4.5\\ 
			$g$  & Network coupling coefficient & N/A & 4\\ 
			$K$  & Axonal conduction speed & N/A & 15\\ 
			\bottomrule
		\end{tabular}
		\vspace{0.1cm}
		\caption{
			\textit{Details for the Wilson-Cowan neural mass model parameters. Each row describes the parameter symbol (left column), a quick description of its biological reference (middle left column) and information concerning the template value and bounds (if applicable) used in each experiment with the following format: default value; [lower bound, upper bound]. N/A refers to instances were the parameter was not included in the model. Note that the displayed noise variances ($\sigma_E$,$\sigma_I$) are on the log scale base 10.}
		}
		\label{table:1}
	\end{table}
	\subsection{Experiment 2 Connectivity}
	The connectivity matrices were designed to mimic properties of empirical human connectomes, including an inverse relationship between connection strength and tract length. A common connectivity matrix was drawn from a beta distribution and then symmetrized and ordered by node degree. Node $n_0$ (highest weighted degree) was systematically manipulated by linearly scaling its input weights, producing 20 synthetic datasets with increasing $S$. This manipulation is illustrated in Figure 5--figure supplement~\ref{figsupp:connectivity}.
	
	\subsection{Supplemental Figures}

\end{document}